\documentclass[twocolumn,showpacs,prl]{revtex4}
\usepackage{amsmath}
\usepackage{epsf}

\newcommand{\beq}{\begin{equation}}
\newcommand{\eeq}{\end{equation}}
\newcommand{\bea}{\begin{eqnarray}}
\newcommand{\eea}{\end{eqnarray}}

\begin{document}

\title{A test of the $g-$ology model for one-dimensional interacting
  Fermi systems} 
\author{Andrey V. Chubukov$^1$, Dmitrii
  L. Maslov$^2$ and Fabian H.L. Essler$^3$} 

\begin{abstract}
Bosonization predicts that the specific heat, $C(T)$, of a
one-dimensional interacting Fermi system is a sum of the
specific heats of  free 
 collective charge and spin excitations, plus the term with
 the running backscattering amplitude which flows to zero logarithmically
 with decreasing $T$.
We verify whether this result is reproduced in the $g-$ology model. 
Of specific interest are the anomalous terms in $C(T)$  that depend on the bare backscattering amplitude. We show that these terms can be incorporated into a
renormalized spin velocity. We do this by proving the equivalence
of the results for $C(T)$ obtained within
 the $g$-ology model and by 
 bosonization with velocities obtained by the 
numerical solution of the Bethe-ansatz equations for the 
Hubbard model.
\end{abstract}

\affiliation{$^1$Department of Physics, University of
Wisconsin-Madison, 1150 Univ. Ave., Madison, WI 53706-1390, USA\\
$^2$Department of Physics, University of Florida, P.
O. Box 118440, Gainesville, FL 32611-8440, USA\\
$^2$ The Rudolf Peierls Centre for Theoretical Physics,\\
University of Oxford, 1 Keble Road Oxford, OX1 3NP, UK
}
\maketitle

One-dimensional interacting fermionic systems are believed to be
described by an effective low-energy theory: the $g$-ology 
model~\cite{solyom,emery,giamarchi_book}.
This model involves a small number of interaction vertices
describing  small momentum scattering of fermions near the same and
opposite Fermi points ($g_4$ and $g_2$, correspondingly) and $2k_F$ scattering
 $g_1$ [for fermions on a lattice, there is also
an Umklapp vertex ($g_3$)]. The effective vertices are, in
principle, obtained by integrating out high-energy fermions in
microscopic models, e.g., the Hubbard model. To first order in $U$, 
$g_i = U/(2\pi v_F)$; beyond first order, all $g_i$ are different,
 and each of them is represented by a   series in $U$.

A powerful way to treat the $g$-ology model is bosonization, which transforms 
interacting fermions into the collective bosonic excitations in the
charge and spin channels~\cite{schulz95,giamarchi_book}.
For the case of a repulsive interaction between original
fermions, considered in this paper, the bosonization shows that
the charge sector is a free Gaussian theory, while the spin sector
becomes asymptotically free in the low-energy limit, when the coupling
of the marginally irrelevant process ($2k_F$ scattering of fermions of
opposite spins) flows to zero~\cite{kondo,cardy,lukyanov,ae} 
As a consequence,  bosonization predicts that  at the lowest
temperatures the specific heat of interacting 1D fermions is the
same as the specific heat of two systems of acoustic 1D phonons,
i.e.,
 \beq C(T) = \frac{\pi T}{3} \left(\frac{1}{v_{\rho}} +
\frac{1}{v_\sigma}\right), \label{1} \eeq 
 where $v_{\rho}$ and
$v_{\sigma}$ are the charge and spin velocities, correspondingly.
Eq.~(\ref{1}) was verified by a weak-coupling renormalization group
(RG) treatment of the original fermionic model, in which all
non-logarithmic corrections to the couplings were neglected~
\cite{solyom,schulz95,giamarchi_book,maslov_review}.
However, it has never been proven explicitly that the 
bosonization result in Eq.~(\ref{1}) is valid to all orders in the
interaction. 

Recent results call for a further study of the validity of Eq.~(\ref{1})
beyond the weak-coupling limit. In particular, 
two of us have obtained~\cite{cm1d}
the specific heat of interacting 1D fermions to order $g_i^3$ and
found  that the effect of $2k_F$ scattering in the spin channel is
more involved than it had been previously thought  -- in addition to terms that
depend on the running coupling $g_1 (T)$,
  the free energy also contains terms
that depend on the bare coupling $g_1$  (see Eq. (\ref{8}) below).

>From the  field-theoretical point of view, these terms should
regarded as {\it anomalies}, i.e., they can be viewed equivalently
either as low- or high-energy contributions.These 
anomalous terms cannot be simply absorbed into the Gaussian part
of the bosonic Hamiltionian, and it is not a priori clear whether
such terms can be incorporated into the renormalized charge and
spin velocities.

The goal of this paper is to prove that the answer 
 to the question formulated above 
 is affirmative, at least within perturbation theory. We show this by comparing the specific heat
obtained from second-order perturbation theory in the $g-$ology model, 
calculated in Ref.~\onlinecite{cm1d}, with Eq. (\ref{1}), where the 
 velocities are obtained from the Bethe ansatz solution of the
Hubbard model \cite{book}.

Our starting point is the Hamiltonian of the 1D Hubbard model
 \beq {\cal H} = \sum_{k,\sigma}
(\epsilon_k - \mu) c^\dagger_{k,\sigma} c^{}_{k,\sigma} + U
\frac{1}{N} \sum_{k,q,l} c^\dagger_{k, \uparrow} c^\dagger_{q,
\downarrow} c^{}_{\downarrow, k+l} c^{}_{\uparrow, q-l}, \label{2}
\eeq where $\epsilon_k = -2t \cos k$ and 
 $\mu=-2t\cos k_F $ is the chemical
potential (the lattice spacing is set to unity). For $U=0$ the Fermi
velocity is related to the Fermi momentum $k_F$ by $v_F=2t\sin k_F$.

Away from
 half-filling, umklapp scattering
 requires collisions of more than two fermions and is therefore neglected 
in our treatment.
The spin and charge velocities can be obtained by solving certain
linear integral equations obtained from the Bethe ansatz solution of
the Hubbard model \cite{book}. 
 At strong coupling, the 
spin velocity is $v_s = (\pi/2) (4t^2/U) (1 - sin{2k_F}/
2 k_F)$~\cite{coll}. The weak coupling limit has been studied in \cite{frahm}.
 It is possible to derive a small $U$
expansion for the spin and charge velocities 
analytically by solving the integral
equations drived in Ref. [\onlinecite{frahm}] by Wiener-Hopf
methods. As we are interested in the ${\cal   O}(U^2)$ terms of the
expansions, the resulting calculations are 
somewhat involved. We therefore have solved the integral equations
numerically for small U and found that the results are well fit by the
series 
\begin{subequations}
\bea
&& v_{\rho}\simeq v_F \left(1 + \frac{U}{2\pi v_F} +c \left(
\frac{U}{2\pi v_F}\right)^2\right) \label{3a} \\ 
&& v_{\sigma}\simeq v_F \left(1 - \frac{U}{2\pi v_F} + c' \left(
\frac{U}{2\pi v_F}\right)^2\right). \label{3b} \eea
\end{subequations} 
Based on the analytic result for the spin velocity at half-filling
\cite{book} and the fact that the only marginally irrelevant operator
(the interaction of spin currents) is the same at and below
half-filling we do not expect logarithmic terms to appear in these
expansions. We have determined the coefficients $c$ and $c'$ for
a number of different densities. The results are shown in Table
\ref{tab:coeffs} 
\begin{table}
\begin{center}
\begin{tabular}{|l|l|l|l|l|l|l|l|l|}
\hline
$n$&0.1&0.2&0.3&0.4&0.5&0.6&0.7&0.8\\
\hline
$c$& -0.51&-0.56&-0.64&-0.74&-0.87&-1.01&-1.16&-1.31\\
\hline
$c'$& 0.49&0.44&0.36&0.26&0.13&-0.01&-0.16&-0.31\\
\hline
\end{tabular}
\caption{Coefficients $c$ and $c'$ for different densities in the 1D
  Hubbard model. The density $n = 2k_F/\pi$; 
$n=1$ corresponds to half-filling.}
\label{tab:coeffs}
\end{center}
\end{table}
As we will see below, we will only need the difference 
$c'-c$ to verify the validity of Eq. (\ref{1}).  In all cases we find
that within the numerical accuracy of our computation 
\begin{equation}
c'-c=1.
\label{BA}
\end{equation}
The deviation from 1 is less than one percent in all cases.  
The spin and charge velocities can alternatively be calculated in
perturbation  theory for the $g-$ology model~\cite{cm1d,matveev}. 
An explicit calculation of $C(T)$ within this model is somewhat
involved 
 as the low-energy $g-$ology model contains  two momentum cutoffs, $\Lambda_f$ 
 and $\Lambda_b$, 
constraining the
integration over the fermionic dispersion  and
over the momentum transfers near $2k_F$, respectively. [The
$g$-ology model is only valid when $\Lambda_{\rm{ f}}  >
\Lambda_{\rm{ b}} $, i.e., when the interaction vanishes at the
cutoff set by the dispersion.] 
Some terms in $C(T)$ are cutoff-independent while some depend
logarithmically on the ratio $\Lambda_{\rm{f}}/\Lambda_{{\rm b}}$.
Fortunately, at least to second  order in $g$, all cutoff-dependent
renormalizations can be absorbed
   into the renormalized backscattering amplitude
    ${\tilde g}_1 = g_1 - 2g^2_1 \log\left({\Lambda_{\rm{ f}} /\Lambda_{\rm{ b}} }\right)$,
      so that the specific heat is expressed in terms of $g_4$, $g_2$, and ${\tilde g_1}$ without any explicit dependence on the cutoffs~\cite{cm1d}.
 To second order in $g$, $C(T)$ is given by
\begin{widetext}
\begin{equation}
C(T)=\frac{2\pi T}{3v_{F}}~\left[ 1+  \left(
{\tilde{g}}_{1}-g_{4}\right) +  \left(
{\tilde{g}}_{1}-g_{4}\right)^2  + g^2_4 + \left(g_{2}
-\frac{1}{2}{\tilde{g}}_{1}\right)^2 + \frac{3}{4}
\tilde{g}_{1}^{2} + O(g^3) \right],  \label{4}
\end{equation}
\end{widetext}
where all vertices are measured in the units of $2\pi v_F$. 

To compare Eq.~(\ref{4}) with Eqs. (\ref{1})
 where
  $v_{\rho}$ and $v_{\sigma}$ given by
Eqs.~(\ref{3a},\ref{3b}), we first note that 
 the $g$-ology model can be bosonized by expressing
 the operators of right- and
left-moving fermions, $R_{\alpha }$ and $L_{\alpha }$ ($\alpha =\uparrow ,\downarrow$)
as
\begin{equation}
R_{\alpha }(x),L_{\alpha }(x)=\frac{1}{\sqrt{2\pi b}} \exp \left[ \pm
i\left( \phi _{\alpha }(x)\mp \theta _{\alpha }(x)\right) \right] ,
\label{5}
\end{equation}
where $b$ is a short-distance cutoff related to the fermionic
momentum cutoff ($\Lambda_{\rm{f}}$) of the $g-$ology model. Under
bosonization, the terms in
 the fermionic Hamiltonian  parameterized by
the couplings $g_{4}$ and $g_{2}$ are mapped onto the free,
Gaussian part of the bosonized Hamiltonian.  The $2k_F$
 term, 
parameterized by  $g_1$, leads to non-linear, cosine
terms in the bosonic Hamiltonian, which give rise to interactions
in the spin channel.

To first order in ${\tilde g}_1$, backscattering just renormalizes
the prefactors in the Gaussian part of the bosonized Hamiltonian~\cite{starykh_maslov,capponi}, 
 so that the $g-$ology model can be reduced to
 a gas of free acoustic bosons with $H_{G}=H_{G}^{\left( \rho
\right) }+H_{G}^{\left( \sigma \right)}$, where
\begin{widetext}
\begin{eqnarray}
&& H_{\mathrm{G}}^{\left( \rho \right) } = \frac{1}{2}\int
dx\left( 1+2 g_{4} + 2 g_{2} - 2{\tilde g}_1\right)
\left( \partial _{x}\phi _{\rho}\right) ^{2}+\left(
1+2 g_{4} -2 g_{2}\right) \left(
\partial _{x}\theta _{\rho}\right) ^{2}, \nonumber \\
&& H_{\mathrm{G}}^{\left( \sigma \right) } = \frac{1}{2}\int
dx\left( 1-2 {\tilde g}_{1}\right)
\left( \partial _{x}\phi _{\sigma}\right) ^{2}+ \left(
\partial _{x}\theta _{\sigma}\right) ^{2},
\label{6}
\end{eqnarray}
\end{widetext}
and the charge and spin bosons are defined as $\phi _{\rho ,\sigma }=\left(
\phi _{\uparrow }\pm \phi _{\downarrow }\right) /\sqrt{2}$ and $\theta
_{\rho ,\sigma }=\left( \theta _{\uparrow }\pm \theta _{\downarrow }\right) /%
\sqrt{2}$.

If this were the only effect of backscattering, the specific heat
would be given by Eq. (\ref{1}) with the effective spin and charge
velocities ${\tilde v}_\sigma$ and ${\tilde v}_\rho$ read off from Eq. (\ref{6}):
\begin{eqnarray}
&&{\tilde v}_{\rho }^{2}\!=\!v^2_F \left((1+2g_{4} -{\tilde g}_1)^{2}\!-\!
(2 g_{2} -{\tilde g}_{1})^{2}\right),  \notag \\
&&{\tilde v}_{\sigma}^{2}\!=\!v^2_F \left((1-{\tilde g}_{1})^{2}\!-\!{\tilde g}_{1}^{2}\right)  \label{7}
\end{eqnarray}
Re-expressing Eq.(\ref{4}) in terms of ${\tilde v}_\sigma$ and
${\tilde v}_\rho$, we obtain \beq
 C(T) = \frac{\pi T}{3} \left(\frac{1}{{\tilde v}_{\rho}} +
\frac{1}{{\tilde v}_{\sigma}}\right) + \frac{\pi T}{3 v_F} {\tilde
g}_{1}^2 + O({\tilde g}^3_1).
\label{8}
 \eeq
We see that this expression differs from Eq. (\ref{1}). This is
not surprising because the effect of
backscattering cannot be simply absorbed into the Gaussian part of
the bosonized Hamiltonian,  beyond the first order in $g_1$.

The issue therefore is whether the extra ${\tilde g}^2_1$ term in Eq. (\ref{8})   can be 
 absorbed into  renormalization of the velocities ${\tilde v} \rightarrow v$, so that the
specific heat is still given by Eq. (\ref{1}) with the 
 renormalized velocities $v_\rho$ and $v_\sigma$. 
 A simple extension of the previous analysis to a non-SU(2) symmetric 
case shows  that, beyond the first order,
 $2k_F$ scattering contributes only to the spin part of the bosonized Hamiltonian.
 The real issue then is renormalization of the spin velocity 
${\tilde v}_\sigma \rightarrow v_\sigma$.
   The charge velocity given by Eq.(\ref{7}) must
 be the same as the exact one, i.e., ${\tilde v}_\rho = v_\rho$.

A straightforward way to check this is to compare Eqs. (\ref{1})
and (\ref{8})
 with Eq. (\ref{3a},\ref{3b}).  Quite generally,
  one can write  $v_\sigma = {\tilde v}_\sigma + a v_F {\tilde g}^2_1$, where $a$ is a dimensionless constant.
   Only if $a=-1$,
 the extra ${\tilde g}^2_1$ term in Eq.(\ref{8}) can be
 absorbed into $v_\sigma$, i.e.,  Eq. (\ref{8}) can be cast into Eq.
 (\ref{1}).
 Using Eq.~(\ref{7}), we obtain
 \begin{subequations}
\begin{eqnarray}
&&{v}_{\rho } = v_F \left(1+2g_{4} -{\tilde g}_1 - \frac{1}{2}
(2 g_{2} -{\tilde g}_{1})^{2}\right),  \label{9a} \\
&& v_{\sigma}\!=\!v_F \left(1-{\tilde g}_{1} - \frac{1-2a}{2}
{\tilde g}^2_{1}\right) \label{9b}
\end{eqnarray}
\end{subequations}
These two expressions should be the same as Eqs.~(\ref{3a},\ref{3b}).

\begin{figure}[tbp]
\begin{center}
\epsfxsize=1.0\columnwidth \epsffile{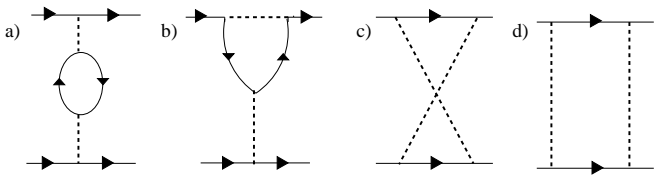}
\end{center}
\caption{One-loop diagrams for the interaction vertices. In this
paper, we need only
 renormalizations coming from the high-energy scales. The 
 high-energy renormalizations of
 $g_4$ and $g_1$ come only from the Cooper diagram d).}
\label{fig:rg}
\end{figure}

To compare Eqs.~(\ref{3a},\ref{3b}) and (\ref{9a},\ref{9b}),
 we need to evaluate renormalizations of $g_4$ and $g_{1}$
 to second order in $U$ and to select the contributions which comes from high energies.  By construction, such contributions are absorbed into the bare couplings of the $g-$ology model.

There are four second-order diagrams for the $g_4$ amplitude
 to order $U^2$ (see Fig. 1). For a constant $U$, diagrams a) and b) cancell each other. Diagram c) contains the polarization bubble
 $\Pi (0) = (1/2\pi)^2 \int d k d \omega/(i\omega - \epsilon_k + \mu)^2$.
 Renormalizations due to $\Pi (0)$ are within the low-energy theory, as
 one can evaluate $\Pi(0)$ in such a way that the result $\Pi(0) = -1/(\pi
 v_F)$ is determined entirely by the states near the Fermi energy
\cite{cm1d}. The remaining diagram d) describes renormalization in
the Cooper channel. Up to a prefactor, it is given by
 \begin{equation}
 \int d k d \omega \frac{1}{(i\omega - \epsilon_{k+k_F}+\mu)(i\omega +
 \epsilon_{k-k_F}-\mu)}.
\label{10_a} \eeq
The momentum integration in Eq.~(\ref{10_a}) is not
confined to the Fermi surface, i.e., this diagram does contribute 
 to high-energy renormalization of $g_4$.
 In the Hubbard model, the momentum integration is limited
from above by the Brillouin zone. The lower limit $\Lambda_f$ can be safely set to zero as the integral
 is infrared-finite, i.e.,  $\int dk = \int _{-\pi}^{\pi} dk $. Integrating over $\omega$
  and then over $k$, we  obtain 
\beq 
g_4 = \frac{U}{2\pi
v_F} \left(1 - \frac{U}{2\pi v_F}\right). \label{10} 
\eeq 
Note that the only depence on the density is through $v_F = 2t \sin k_F$.
We verified that a model of fermions in a continuum with dispersion
$k^2/2m$ gives the same result.

Renormalization of $g_1$ is more 
involved
 because $g_1$ is a running
coupling, and the second-order result depends logarithmically on the
upper cutoff of the low-energy theory, which is the lower limit of
the integration for ``high-energy'' renormalization.
 The diagrams are the same as in Fig. 1. As for $g_4$, diagrams a) and b) cancel each other while diagram c) contains
 $\Pi(0)$. Therefore, only Cooper diagram d) contributes to the ``high-energy'' renormalization.
 Evaluating Cooper diagram for backscattering, we obtain
\beq
g_1 = (U/2\pi v_F) \left (1 - L ~\frac{U}{2 \pi v_F}\right),
\label{s_1}
\eeq
 where 
\beq 
L = \frac{v_F}{2 \pi}
 {\cal P} \int_{-\pi}^{\pi} dk \int_{-\infty}^\infty  
 d\omega \frac{1}{\omega^2 + \left(\epsilon_{k}-\mu\right)^2}
\label{10_aa}
 \eeq
 and 
  The
symbol ${\cal P} \int$ here implies that momentum integral does
not include the regions near the Fermi points
of width $2\Lambda_{{\rm f}}$. Integrating over
frequency and
then over momentum, we obtain
\beq 
L = 2 \log{\frac{2 \sin{k_F}}{\Lambda_{\rm{ f}}}} \label{10_b} 
\eeq
 Substituting Eq.~(\ref{10_b}) into Eq.~(\ref{s_1}), we then obtain
 \beq
 g_1 = \frac{U}{2\pi v_F} \left( 1 - \frac{U}{\pi v_F} \log 
{\frac{2 \sin{k_F}}{\Lambda_{\rm{ f}}}}\right)
\label{10_cc}
\eeq
In addition, $g_1$ is renormalized {\it within} the low-energy g-ology model.
 This renormalization depends logarithmically on the ratio of the bosonic and fermionic cutoff of the $g-$ology model:
${\tilde g}_1 = g_1 -  2 g^2_1 \log\left(\Lambda_{\rm{ f}}
/\Lambda_{\rm{ b}} \right)$ \cite{cm1d}. Adding up this result
with Eq.~(\ref{10_c}) we find that the combination of the high-energy
and low-energy renormalizations just replaces $\Lambda_{\rm{ f}} $
by $\Lambda_{\rm{ b}} $ under the logarithm, i.e.,
  \beq {\tilde g}_1
= \frac{U}{2\pi v_F} \left( 1 - \frac{U}{\pi v_F}
\log{\frac{2 \sin{k_F}}{\Lambda_{\rm{ b}}}}\right). \label{10_c} \eeq

Alternatively, one can obtain the full renormalization of
$g_1$ by excluding the regions of width $\Lambda_{\rm{b}}$ near
$\pm 2k_F$ from the integration over the momentum transfer $k$.
This gives the same result as in Eq.~(\ref{10_c}).

The value of $\Lambda_{\rm{ b}} $ is unknown: as we said earlier, the
$g-$ology model {\it assumes} that the low-energy properties of the
original system of 1D fermions with a short-range interaction  are the
same as in the model where interactions are artificially restricted to
narrow regions of momentum transfers either near zero 
(for $g_4$ and $g_2$)  or
$2k_F$ (for $g_1$).
 We can only realistically expect that
$ v_F\Lambda_{\rm{ b}} $ is 
 substantially smaller than 
a half of the fermionic bandwidth $W/2=2t$.
 Still, we have two pairs of equations to compare [Eqs.~
  (\ref{3a},\ref{3b}) and (\ref{9a},\ref{9b})] and two unknown
parameters: $\Lambda_{\rm{ b}} $ and $a$. Solving for the unknowns,
we obtain 
\begin{eqnarray}
\Lambda_b &=& 2 \sin{k_F}~\exp\Bigl(-\frac{5}{4}-\frac{c}{2}\Bigr)\ ,\nonumber\\
a&=&c'-c-2.
\end{eqnarray}
 By virtue of Eq.~(\ref{BA}), we conclude that $a=-1$, which is precisely
the value of $a$ one needs to cast Eq. (\ref{8}) into Eq. (\ref{1}).
This, we believe, is a ``numerical proof'' of the statement that at
very low temperatures the specific heat of 1D interacting fermions is
the same as two system of acoustic phonons with certain spin and
charge velocities.   

We also find that for most of
 densities $v_F \Lambda_b/(2t) = \Lambda_b \sin k_F$
is smaller than one, as it should be, otherwise the $g-$ology model 
cannot be justified. In particular, at  quarter-filling, 
$ \Lambda_b \sin k_F \sim 0.44$.  
 Near half-filling, however, $\Lambda_b \sin k_F$ becomes larger than one,
  which questions the validity of the $g-$ology model.
  Note also that in the opposite limit of small density,
   $\Lambda_b$ goes to zero as it indeed should as at vanishing $k_F$ the linearized dispersion no longer holds.  
 
To conclude, in this paper we obtained expressions for spin and charge
velocities for interacting 1D fermions in terms of the couplings of 
the $g$-ology model. These can be related to the microscopic
parameters of the 1D Hubbard model via a comparison to weak coupling
expansions of the velocities obtained from the Bethe ansatz solution.
Using these results, we have shown that all terms in the specific heat
in the $g-$ology model that do not flow under RG  are absorbed into
the specific heat of two free gases of massless bosons.
 As a result, the full specific heat of a system of
interacting fermions in 1D is a sum of the specific heats  of two
free massless Bose gases of charge and spin excitations, and the  true
interaction term, which contains the running backscattering
amplitude and logarithmically flows to zero with decreasing $T$. 

We acknowledge helpful discussions with
I. L. Aleiner, H. Frahm, T. Giamarchi, L. I. Glazman,
K. B. Efetov, A. W. W. Ludwig, K. A. Matveev, A. Melikyan, A.M. Tsvelik, and
support from NSF-DMR 0604406 (A.V. Ch), NSF-DMR 0308377 (D. L. M.)
and the EPSRC under grant GR/R83712/01 (FHLE).

\end{document}